\documentclass[aps,prd,preprint,unsortedaddress,amsmath,nofootinbib]{revtex4-1}
\usepackage{graphicx}
\usepackage{dcolumn}
\usepackage{bm}
\usepackage{mathrsfs}
\usepackage{amsfonts}
\usepackage{slashed}
\usepackage{natbib}

\begin{document}
\bibliographystyle{plainnat}
\bibliographystyle{abbrvnat}
\bibliographystyle{unsrtnat}
\title{Electromagnetism in Quark-Antiquark Bound States}
\author{R. Ling}
\email{raimund.ling@gmail.com}
\affiliation{Department of Physics, Nanjing University, Nanjing 210093, China}
\author{B. L. Lee}
\affiliation{School of Physics and Technology, Nanjing Normal University, Nanjing 210046, China}
\author{H. S. Zong}
\email{zonghs@chenwang.nju.edu.cn}
\affiliation{Department of Physics, Nanjing University, Nanjing 210093, China}
\affiliation{Joint Center for Particle, Nuclear Physics and Cosmology, Nanjing 210093, China}
\author{J. L. Ping}
\affiliation{School of Physics and Technology, Nanjing Normal University, Nanjing 210046, China}
\begin{abstract}
Non-perturbative proof is presented of a unique version of Goldstone theorem, that electromagnetism contributes to the masses of spinless particles of quark-antiquark bound states in the form of a commutator of the quark electric matrix and a coefficient factor which may be expressed simply in terms of a three-body Bethe-Salpeter amplitude, in the case where the Lagrangian conserves an approximate $SU(3)$ chiral symmetry. The coupled Bethe-Salpeter equation and Dyson-Schwinger equation in the ladder-rainbow approximation are layed out with every ingredient renormalized, which are applied to the problem of evaluating the mass differences of some charged and neutral pseudoscalar and vector mesons. The numerical results yield satisfactory agreement with experimental observations for the mass spectrum and meson decay constants. Estimates of the light quark masses are also given in the Bethe-Salpeter formalism.
\end{abstract}
\maketitle

\section{Introduction}
Although it is much weaker than the strong interaction at realistic energy scales in nuclear or elementary particle physics on general grounds, the electromagnetic interaction plays a significant role in accounting for the emperical observation of the mass differences between charged and neutral mesons and within an octet or a decuplet of baryons. To the knowledge of quantum chromodynamics, the modern theory of the strong interaction, quarks that are the elementary degrees of freedom of the theory are never isolated; a study of their propertities like the quark masses must come from the information of the observation of the bound states of quarks (and antiquarks). To make it concrete, in utilizing the experimental data of the masses of particles that are sensitive to the masses of the three fairly light quarks u, d and s, such as light mesons like the isotopic doublet kaon that behaves as a bound state of an $\bar{s}$ antiquark and a u or d quark, or ground-state nucleons like the proton and neutron that are approximately regarded as colour-neutral bound states of three quarks of types u and d, we must take into consideration of the electromagnetic interaction to offer an estimate of the scales of the masses of those confined quarks.

In the past the electromagnetic self-energies of the ground-state mesons have been examined successfully to leading order in the electromagnetic interaction in chiral perturbation theory.\citep{dashen69,lan73} Non-perturbative methods like the Bethe-Salpeter calculation have also been used to determine the electromagnetic as well as isospin mass splittings for the pseudoscalar mesons as well as the vector and scalar mesons.\citep{jain93} There seems to be a recent revival of attention paid to the role of the electromagnetic correction in the computation of hadron observables to high precision by lattice stimulation\citep{blum2007,blum2010}. The purpose of the present paper is to derive some general results valid to all orders in perturbation theory which should provide the essential bridge between the approximate symmetry of the underlying theory and the realization of the physical states of the theory which manifest themselves in something like the spectrum of composite particles of interest, and give a unified and definite description of light mesons like ground-state pseudoscalar and vector mesons and gain numerical results to compare with experimental data. 

Sec. II starts with the derivation of the axial-vector Ward identity in the general sense that the electromagnetic interaction term is added to the Lagrangian density of quantum chromodynamics, which makes a connection between the approximate invariance of the Lagrangian density with respect to the $SU(3)$ chiral transformation and physical quantities with radiative corrections taken into account. Then by clarification of the structure of Feynmann amplitudes between vacuum states shown in the generalized axial-vector Ward identity, we understand that the existence of poles arises from pseudo-Goldstone bosons which can be simultaneously interpreted as bound states of elementary particles with their own fields in the Lagrangian. It follows that an exact non-perturbative mass formula is obtained for the pseudoscalar octet, with constant coefficients which may be expressed in terms of two-body or three-body Bethe-Salpeter amplitudes, the matrix elements which are equivalent to the sums of all Feynmann diagrams with incoming lines on the mass shell corresponding to the one-pseudo-Goldstone-boson states, and lines off the mass shell corresponding to the operators of elementary quark (and photon) fields. We will see in this section the peculiar though understandable features in the case of the combination of two fundamental theories of the standard model.

As a complement to Sec. II, Sec. III deals with mesons of quark-antiquark pairs which appear as solutions of the ordinary Bethe-Salpeter equation. With renormalization considered explicitly, the electromagnetic interaction kernel is incorporated into the system by being added to the usual strong interaction kernel, and the Bethe-Salpeter equation is solved consistently with the Dyson-Schwinger equation for the quark propagators, which is necessary to demonstrate the Goldstone nature of the pion in the limit of zero quark masses. The reason why we restrict ourselves there in the two-body system is that the approach we used is simple and can be easily extended to the widest range of the spectrum of mesons of arbitrary type.

Our numerical results are presented in Sec. IV, where it is assumed that the first one or two terms in the Chebyshev expansion of the Bethe-Salpeter amplitudes give the dominant contributions, though a full Lorentz-component calculation is carried out always. We employ an existed \emph{Ansatz} of the strong effective interaction and solve in a straightforward way for the numerical solution of the Dyson-Schwinger equation that will be substituted into the Bethe-Salpeter equation, by analytic continuation of the effective interaction function to the complex plane, without compromising on some analytic structure that may impair the subsequent explorations of meson decays and form factors.

For the greater part of this review apart from numerical calculation, we use the spacetime metric $\eta_{\mu\nu}$ with diagonal elements $\eta_{11}=\eta_{22}=\eta_{33}=1$, $\eta_{00}=-1$, and Dirac matrices $\gamma_{\mu}$ defined so that $\{\gamma_{\mu},\gamma_{\nu}\}=2\eta_{\mu\nu}$, also $\gamma_5=i\gamma_0\gamma_1\gamma_2\gamma_3$, and $\beta=i\gamma^0$.

\section{Non-Perturbative Elaboration}
As is known to all, in the approximation that the u, d and s quarks are massless, the Lagrangian of quantum chromodynamics is invariant with respect to an $SU(3)$ transformation which implies the existence of a conserved axial-vector current
\begin{equation}
J^{\mu}_{a}=i\bar{q}\gamma^{\mu}\gamma_{5}t_{a}q,
\end{equation}
where q is the quark triplet,
\begin{equation}
q\equiv\begin{pmatrix}
u\\
d\\
s
\end{pmatrix},
\end{equation}
and $t_a$ may be Gell-Mann--Ne'eman matrices $\lambda_a$ themselves or suitable linear combinations of them.

In order to see the relation between the quark propagator and the axial-vector vertex of the theory, let us consider the momentum-space Green's function for the axial-vector current $J_{a}^{\mu}(x)$, together with a Heisenberg-picture quark field $q_n(y)$ and its covariant adjoint $\bar{q}_{m}(z)$. We define the axial-vector vertex function $\Gamma_{a}^{\mu}$ by
\begin{align}
\int\mathrm{d^4}x\mathrm{d^4}y\mathrm{d^4}z e^{-iq_1\cdot x}e^{-iq_2\cdot y}e^{+iq_3\cdot z}\langle T\{ J_{a}^{\mu}(x)q_n(y)\bar{q}_{m}(z)\}\rangle_{\mathrm{VAC}}\nonumber\\
\equiv -i(2\pi)^{4}S'_{n}(q_2)[\Gamma_{a}^{\mu}]_{nm}(q_2,q_3)S'_{m}(q_3)\delta^{4}(q_1+q_2-q_3),
\end{align}
where
\begin{equation}
-i(2\pi)^{4}S'_{n}(q_2)\delta^{4}(q_2-q_3)\equiv\int\mathrm{d^4}y\mathrm{d^4}z e^{-iq_2\cdot y}e^{+iq_3\cdot z}\langle T\{ q_n(y)\bar{q}_{n}(z)\}\rangle_{\mathrm{VAC}}
\end{equation}
is the complete quark propagator of a particular flavour $n$. By use of the conservation condition $\partial_{\mu}J_{a}^{\mu}=0$ and the commutation relation between the time-component of the current $J_{a}^{0}$ and the quark fields $q$, we can derive the celebrated axial-vector Ward identity
\begin{equation}
(q_3-q_2)_{\mu}\Gamma_{a}^{\mu}(q_2,q_3)=i\gamma_{5}t_{a}S'^{-1}(q_3)+iS'^{-1}(q_2)\gamma_{5}t_{a},
\end{equation}
from which the nature of the cancellations among the radiative corrections to the quark propagators and the axial-vector vertex is manifest.

In the real world the $SU(3)$ symmetry of quantum chromodynamics is broken by the quark mass term and the axial-vector current is not exactly conserved; its divergence that can be obtained from the Euler-Lagrange equation is given by
\begin{equation}
\partial_{\mu}J_{a}^{\mu}=i\bar{q}\gamma_{5}\{M_q,t_{a}\}q,
\end{equation}
where the quark mass matrix $M_q$ is a diagonal matrix with non-zero elements $m_u$, $m_d$ and $m_s$.
In this case the axial-vector Ward identity yields the form
\begin{equation}
i(q_3-q_2)_{\mu}\Gamma_{a}^{\mu}(q_2,q_3)=-\gamma_{5}t_{a}S'^{-1}(q_3)-S'^{-1}(q_2)\gamma_{5}t_{a}+\{M_q,\Gamma_{a}(q_2,q_3)\},
\end{equation}
where the pseudoscalar vertex function $\Gamma_{a}$ is defined by
\begin{align}
\int\mathrm{d^4}x\mathrm{d^4}y\mathrm{d^4}z e^{-iq_1\cdot x}e^{-iq_2\cdot y}e^{+iq_3\cdot z}\langle T\{ \phi_{a}(x)q_n(y)\bar{q}_{m}(z)\}\rangle_{\mathrm{VAC}}\nonumber\\
\equiv -i(2\pi)^{4}S'_{n}(q_2)[\Gamma_{a}]_{nm}(q_2,q_3)S'_{m}(q_3)\delta^{4}(q_1+q_2-q_3),
\end{align}
with the pseudoscalar current $\phi_{a}=i\bar{q}\gamma_{5}t_{a}q$.

In a similar fashion we take into account another small correction, the effect of electromagnetism. It makes an extra contribution, by adding the electromagnetic interaction term to the Lagrangian of quantum chromodynamics, to the divergence of the axial-vector current
\begin{equation}
\partial_{\mu}J_{a}^{\mu}=i\bar{q}\gamma_{5}\{M_q,t_{a}\}q-eA_{\mu}\bar{q}\gamma^{\mu}\gamma_{5}[Q_q,t_{a}]q,
\end{equation}
where $Q_q$ is a diagonal matrix with elements $2/3$, $-1/3$ and $-1/3$.
Now a new electromagnetic pseudoscalar vertex function $\Gamma^{A}_{a}$ which corresponds with the so-called electromagnetic pseudoscalar current $O_a\equiv A_{\mu}J_a^{\mu}=A_{\mu}i\bar{q}\gamma^{\mu}\gamma_{5}t_{a}q$ must be included and is defined by
\begin{align}
\int\mathrm{d^4}x\mathrm{d^4}y\mathrm{d^4}z e^{-iq_1\cdot x}e^{-iq_2\cdot y}e^{+iq_3\cdot z}\langle T\{ O_{a}(x)q_n(y)\bar{q}_{m}(z)\}\rangle_{\mathrm{VAC}}\nonumber\\
\equiv -i(2\pi)^{4}S'_{n}(q_2)[\Gamma^{A}_{a}]_{nm}(q_2,q_3)S'_{m}(q_3)\delta^{4}(q_1+q_2-q_3).
\end{align}
We may eventually give the generalized axial-vector Ward identity in its full form
\begin{equation}
i(q_3-q_2)_{\mu}\Gamma_{a}^{\mu}(q_2,q_3)=\{M_q,\Gamma_{a}(q_2,q_3)\}+ie[Q_q,\Gamma^{A}_{a}(q_2,q_3)]-\gamma_{5}t_{a}S'^{-1}(q_3)-S'^{-1}(q_2)\gamma_{5}t_{a},
\end{equation}
which clarifies the way of the cancellations among the great variety of the radiative corrections to the quark propagators and vertices induced by three kinds of currents all with negative parity and zero baryon number.

To look in a little more detail at the significance of these three currents, let us consider the momentum-space amplitude
\begin{align}
G_{lnm}(q_1,q_2,q_3)&=\int\mathrm{d^4}x\mathrm{d^4}y\mathrm{d^4}z e^{-iq_1\cdot x}e^{-iq_2\cdot y}e^{+iq_3\cdot z}\times\nonumber\\
&\times\langle T\{\mathscr{O}_{l}(x)q_n(y)\bar{q}_{m}(z)\}\rangle_{\mathrm{VAC}},
\end{align}
where $\mathscr{O}_{l}(x)$ can be any of the three currents appearing in the generalized axial-vector Ward identity (11). From Lorentz invariance and parity conservation we know that generally $\mathscr{O}_{l}$ has a non-zero matrix element between the vacuum and a one-(pseudo-)Goldstone-boson state $|B,\mathbf{q}\rangle$, which also has a non-vanishing matrix element with the state $q_n\bar{q}_m|\mathrm{VAC}\rangle$. Then according to the usual rules of polology\citep{lurie65}, $G$ has a pole at $q_1^2=-m^2$, where m is the mass of the one-particle state, and the residue at this pole is given by\footnote{Here we adopt the more usual normalization convention $\langle\mathbf{q}',\sigma'|\mathbf{q},\sigma\rangle=\delta_{\sigma'\sigma}\delta^3(\mathbf{q}'-\mathbf{q})$ with $\mathbf{q}$, $\mathbf{q}'$ the three-mementa, $\sigma$, $\sigma'$ discrete labels for something like the spin three-components of the one-particle states $|\mathbf{q},\sigma\rangle$ and $|\mathbf{q}',\sigma'\rangle$.}
\begin{align}
&G_{lnm}(q_1,q_2,q_3)\to\frac{-2i\sqrt{\mathbf{q}_1^2+m^2}}{q_1^2+m^2-i\epsilon}(2\pi)^3\langle\mathrm{VAC}|\mathscr{O}_l(0)|B,\mathbf{q}_1\rangle\times\nonumber\\
&\times\int\mathrm{d^4}y\mathrm{d^4}z e^{-iq_2\cdot y}e^{+iq_3\cdot z}\langle B,\mathbf{q}_1|T\{q_n(y)\bar{q}_{m}(z)\}|\mathrm{VAC}\rangle.
\end{align}
Conventionally we may define the conjugate Bethe-Salpeter amplitude $\bar{\chi}$ of renormalized fields by
\begin{align}
&\int\mathrm{d^4}y\mathrm{d^4}z e^{+iq_1\cdot(c_1y+c_2z)}e^{-ik\cdot y}e^{+ip\cdot z}\langle B,\mathbf{q}_1|T\{q_{n}(y)\bar{q}_m(z)\}|\mathrm{VAC}\rangle\nonumber\\
&\equiv -(2\pi)^4\delta^4(p-k)\bar{\chi}_{nm}(k;q_1)/[(2\pi)^{3/2}(2\sqrt{\mathbf{q}_1^2+m^2})^{1/2}],
\end{align}
where for quark propagators the incoming relative momentum $k=q_2+c_1q_1$ and outgoing relative momentum $p=q_3-c_2q_1$, with $c_1+c_2=1$.

To be specific, for a (pseudo-)Goldstone boson B of four-momentum $q^{\mu}$, Lorentz invariance and isospin symmetry require the matrix elements of the currents between the vacuum and single-particle states to take the respective forms
\begin{equation}
\langle\mathrm{VAC}|J_a^{\mu}(0)|B_b\rangle=\frac{i\sqrt{2}F_b\delta_{ab}q^{\mu}}{(2\pi)^{3/2}Z_2\sqrt{2q^0}},
\end{equation}
\begin{equation}
\langle\mathrm{VAC}|\phi_a(0)|B_b\rangle=\frac{\sqrt{2}N_b\delta_{ab}}{(2\pi)^{3/2}Z_4\sqrt{2q^0}},
\end{equation}
and also
\begin{equation}
\langle\mathrm{VAC}|O_a(0)|B_b\rangle=\frac{\sqrt{2}R_b\delta_{ab}}{(2\pi)^{3/2}Z_A\sqrt{2q^0}},
\end{equation}
where $F$, $N$ and $R$ are all constant coefficients to be determined, and $Z_2$ is the renormalization constant of the quark fields, $Z_4\equiv Z_2Z_m$ with $Z_m$ the renormalization constant of the quark mass, and $Z_A\equiv Z_2Z_3^{-1/2}$ with $Z_3$ the renormalization constant of the electromagnetic field.\footnote{The renormalization point can be set to be large enough so as to ensure that the renormalization constants are flavour independent.} All the renormalization constants are chosen so that renormalized quantities of the theory preserve the underlying symmetry as in the axial-vector Ward identity (11).\citep{prep68,mrt98}

We can write the (pseudo-)Goldstone boson fields in a real basis as $\pi_a$, with $B_{1/2}\equiv\pi^\pm=(\pi_1\pm i\pi_2)/\sqrt{2}$, $B_3\equiv\pi^0=\pi_3$, $B_{4/5}\equiv K^\pm=(\pi_4\pm i\pi_5)/\sqrt{2}$, $B_6\equiv K^0=(\pi_6+i\pi_7)/\sqrt{2}$, $B_7\equiv\bar{K}^0=(\pi_6-i\pi_7)/\sqrt{2}$, and $B_8\equiv\eta^{0}=\pi_8$. The generators $t_a$ which represent the surviving isotopic $SU(3)$ symmetry are taken in parallelism as $t_{1/2}=(\lambda_1\mp i\lambda_2)/\sqrt{2}$, $t_3=\lambda_3$, $t_{4/5}=(\lambda_4\mp i\lambda_5)/\sqrt{2}$, $t_{6/7}=(\lambda_6\mp i\lambda_7)/\sqrt{2}$, and $t_8=\lambda_8$. In this way the constant factors defined by Eqs. (15)-(17) can be expressed explicitly in terms of the quark fields which are supposed to constitute the corresponding quark-antiquark bound states
\begin{equation}
\langle\mathrm{VAC}|i\bar{q}_{m}(0)\gamma^{\mu}\gamma_{5}q_n(0)|B,\mathbf{q}\rangle=\frac{iF_{mn}^Bq^{\mu}}{(2\pi)^{3/2}Z_2\sqrt{2q^0}},
\end{equation}
\begin{equation}
\langle\mathrm{VAC}|i\bar{q}_{m}(0)\gamma_{5}q_n(0)|B,\mathbf{q}\rangle=\frac{N_{mn}^B}{(2\pi)^{3/2}Z_4\sqrt{2q^0}},
\end{equation}
and also
\begin{equation}
\langle\mathrm{VAC}|A_{\mu}(0)i\bar{q}_{m}(0)\gamma^{\mu}\gamma_{5}q_n(0)|B,\mathbf{q}\rangle=\frac{R_{mn}^B}{(2\pi)^{3/2}Z_A\sqrt{2q^0}}.
\end{equation}

The factor $F$ in the form (18) may be expressed in terms of the renormalized Bethe-Salpeter amplitude $\chi$ that is defined by
\begin{align}
&(2\pi)^4\delta^4(k-p)\chi_{nm}(p;q)/[(2\pi)^{3/2}(2q^0)^{1/2}]\nonumber\\
\equiv\int\mathrm{d^4}y\mathrm{d^4}z& e^{-iq\cdot(c_1y+c_2z)}e^{-ik\cdot y}e^{+ip\cdot z}\langle\mathrm{VAC}|T\{q_{n}(y)\bar{q}_m(z)\}|B,\mathbf{q}\rangle,
\end{align}
where for external fermion lines the incoming shifted momentum $k=q_2-c_1q$ and outgoing shifted momentum $p=q_3+c_2q$, with $c_1+c_2=1$. Note that the relation between the Bethe-Salpeter amplitude and its conjugate is given by 
\begin{equation}
\bar{\chi}(p;q)^T=\beta\chi(-p;q)^{\dagger}\beta.
\end{equation}
Multiplying both sides of Eq. (21) by $iq_{\mu}\gamma^{\mu}\gamma_5$, taking the trace over the colour and spinor indices, integrate over the four-momenta $k$ and $p$ and from the definition (18), one obtains for $q^2=-m_B^2$\citep{jain91}
\begin{equation}
F_B=\frac{Z_2n_cq_{\mu}}{m_B^2}\int\frac{\mathrm{d^4}p}{(2\pi)^4}Tr\{\gamma^{\mu}\gamma_5\chi_B(p;q)\},
\end{equation}
where for colour-neutral bound states $B$ the number of colours $n_c=3$ is factored out explicitly. The constants $F_B$ have an official name of the pseudoscalar meson decay constants, one example of which is $F_{\pi^+}$, the factor involved in the process of pion decay, $\pi^+\to\mu^++\nu_\mu$.

Similarly, by multiplying both sides of Eq. (21) by $i\gamma_5$, taking the trace throughout, and evaluating the integrals over $k$ and $p$ we may get from (19) the expressions of $N_B$ in terms of the Bethe-Salpeter amplitudes\citep{mrt98}
\begin{equation}
N_B=-iZ_4n_c\int\frac{\mathrm{d^4}p}{(2\pi)^4}Tr\{\gamma_5\chi_B(p;q)\},
\end{equation}
with $q$ fixed by $q^2=-m_B^2$.

In order to obtain for $R_B$ a formula we have to define the so-called three-body Bethe-Salpeter amplitude (still denoted by $\chi$) by
\begin{align}
\int&\mathrm{d^4}x\mathrm{d^4}y\mathrm{d^4}z e^{-iq\cdot(c_1x+c_2y+c_3z)}e^{-il\cdot x}e^{-ik\cdot y}e^{+ip\cdot z}\times\nonumber\\
&\times\langle\mathrm{VAC}|T\{A^{\mu}(x)q_{n}(y)\bar{q}_m(z)\}|B,\mathbf{q}\rangle\nonumber\\
\equiv(2\pi)^4&\delta^4(l+k-p)\chi^{\mu}_{nm}(k,p;q)/[(2\pi)^{3/2}(2q^0)^{1/2}],
\end{align}
which gives the sum of all Feynman graphs for emitting a Goldstone boson of four-momentum $q$, with one incoming photon line of relative momentum $l=q_1-c_1q$, one incoming quark line of relative momentum $k=q_2-c_2q$ and one outgoing quark line of relative momentum $p=q_3+c_3q$, with $c_1+c_2+c_3=1$.
Multiply both sides of Eq. (25) by $i\gamma^{\mu}\gamma_5$, take the trace throughout, integrate over the four-momenta $l$, $k$ and $p$ and use Eq. (20), we may obtain finally
\begin{equation}
R_B=-iZ_An_c\int\frac{\mathrm{d^4}k}{(2\pi)^4}\int\frac{\mathrm{d^4}p}{(2\pi)^4}Tr\{\gamma_{\mu}\gamma_5\chi_B^{\mu}(k,p;q)\}.
\end{equation}

Eq. (13) with (14)-(17) substituted in shows the pole structure of the Green's functions associated with the three significant currents $J_a^{\mu}$, $\phi_a$ and $O_a$. Equating the pole terms in the generalized axial-vector Ward identity (11) entails the exact mass formula for each pseudo-Goldstone boson
\begin{equation}
m_B^2F_B=-\{M_q,N_B\}-ie[Q_q,R_B],
\end{equation}
where the constants $F$, $N$ and $R$ are given by Eqs. (23),(24) and (26) in terms of Bethe-Salpeter amplitudes.

Let us now exploit the physical significance of the mass formula (27) at low energy. In the simplest case where the quarks are massless, we consider the matrix element of the time-ordered product of the axial-vector current, the quark field and the covariant adjoint quark field. Substituting Eqs. (18) and (14) into Eq. (13), we get\footnote{In the case where the energies of interest are much lower than the scale of the renormalization point, it is conventional to use the field renormalization prescription that $Z_2=1$.}
\begin{equation}
G^{\mu}_{nm}(q,k,p)\to\frac{-\sqrt{2}F_Bq^{\mu}}{q^2}\bar{\chi}^B_{nm}(p;q)(2\pi)^4\delta^4(p-k).
\end{equation}
Only the term proportional to $\gamma_5$ makes a contribution to the Bethe-Salpeter amplitude as $q^2=0$, so we may write
\begin{equation}
\chi(p;q^2=0)=\gamma_5E(p^2),
\end{equation}
with $E$ a real function of the only scalar variable $p^2$ for real $p^2$. From Eq. (22) we know
\begin{equation}
\bar{\chi}(p;q^2=0)=-\gamma_5E(p^2).
\end{equation}
On the other hand, the inverse of the complete quark propagator $S'^{-1}$ takes the form
\begin{equation}
S'^{-1}(p)=i\gamma_{\mu}p^{\mu}A(p^2)+B(p^2),
\end{equation}
where $A$ and $B$ are coefficient functions of the scalar variable $p^2$. For zero quark masses, there is no distinction of $A$'s and $B$'s among the cases of u, d and s quarks. The non-pole contributions to the Green's function $G^{\mu}$ when contracted with $q_{\mu}$ vanish as $q\to0$, so we may substitute Eq. (28) with $\bar{\chi}$ replaced by (30) into the left-hand side of the axial-vector Ward identity (5), and (31) into the right-hand side, and obtain a formula expression of $E$ in terms of the $A$ and $B$
\begin{equation}
E(p^2)=\frac{2}{F}\frac{B(p^2)}{p^2A^2(p^2)+B(p^2)},
\end{equation}
with a universal constant $F$ for all very soft Goldstone bosons. Then Eq. (24) requires
\begin{equation}
FN=2v,
\end{equation}
where in lowest order the vacuum expectation value of the quark bilinears are all equivalent to an unaltered value $\langle\bar{u}u\rangle_0=\langle\bar{d}d\rangle_0=\langle\bar{s}s\rangle_0\equiv -v/Z_4$, with $v$ given by
\begin{equation}
v=-iZ_4n_c\int\frac{\mathrm{d^4}p}{(2\pi)^4}Tr\{S'(p)\}.
\end{equation}
We see that the quark electric charge matrix $Q_q$ commutes with $t_3$, $t_6$, $t_7$ and $t_8$, so inspection of Eq. (9) shows  that to all orders in quark masses electromagnetic effects give no masses to the neutral pseudo-Goldstone bosons $\pi^0$, $K^0$, $\bar{K}^0$ and $\eta^0$. Also, the electromagnetic part of the divergence of the axial-vector current remains unchanged whether it be $t_1$ or $t_4$ in the limit of zero quark masses, and hence in this case the electromagnetic corrections to the $K^+$ and $\pi^+$ masses are equal. Similar results can be obtained for $\pi^-$ and $K^-$.

Therefore with full effects including electromagnetism taken into account, the mass formulae (27) reads for the pesudoscalar octet to first order in quark masses,\citep{dashen69,lan73,wein77}
\begin{align}
&m_{\pi^{\pm}}^2=2v(m_u+m_d)/F^2+\Delta,\nonumber\\
&m_{\pi^0}^2=2v(m_d+m_u)/F^2,\nonumber\\
&m_{K^{\pm}}^2=2v(m_u+m_s)/F^2+\Delta,\\
&m_{K^0}^2=m_{\bar{K}^0}^2=2v(m_d+m_s)/F^2,\nonumber\\
&m_{\eta^0}^2=2v(4m_s+m_d+m_u)/3F^2,\nonumber
\end{align}
where $\Delta$ is the common electromagnetic correction to the $K^+$ and $\pi^+$ squared masses.

\section{Bethe-Salpeter Equation}
Ordinary mesons like the $\rho^{\pm}$, $J/\psi$ as well as pseudoscalar mesons like $\pi^{\pm}$ are usually interpreted as bound states of quark-antiquark pairs with definite quantum numbers, accompanied by the description of a two-body Bethe-Salpeter amplitude defined as by Eq. (21). This amplitude can be obtained by solving a relativistic integral equation\citep{bethe51,gellmann51}
\begin{equation}
S'^{-1}_{n}(p+c_1q)\chi_{nm}(p;q)S'^{-1}_{m}(p-c_2q)=\int\mathrm{d^4}k K(p,k)\chi_{nm}(k;q)
\end{equation}
with the Dirac and colour indices suppressed, where $K$ is the interaction kernel giving the sum of all Bethe-Salpeter irreducible diagrams and $c_1+c_2=1$. For the sake of simplicity and universality, we from now on confine ourselves to two-body systems, where the electromagnetic effects are incorporated by including photon exchange kernel to the usual Bethe-Salpeter strongly interactive kernel. In this case the integral equation for the complete propagator of renormalized quark fields as in Eq. (31) must be solved consistently, by including electromagnetic corrections to the quark self-energy. This integral equation known as the Dyson-Schwinger equation takes the form
\begin{equation}
S'^{-1}(p)=Z_2S^{-1}(p)+\Sigma(p),
\end{equation}
where the quark flavour indices are suppressed, the inverse of the bare quark propagator of a given flavour $n$ is $S^{-1}_{n}(p)=i\slashed{p}+Z_mm_n$ with the renormalized quark mass $m_n$, and the quark self-energy function $\Sigma$ can be written as a sum of $\Sigma_s$ and $\Sigma_e$, with the contribution from the pure strong interaction
\begin{equation}
i(2\pi)^4\Sigma_s(p)=\frac{4}{3}Z_{1}g^2\int\mathrm{d^4}k\mathfrak{D}'_{\mu\nu}(p-k)\gamma^{\mu}S'(k)\mathfrak{\Gamma}^{\nu}(k,p),
\end{equation}
where $(\lambda_{\alpha}/2)\mathfrak{\Gamma}^{\nu}(k,p)$ is the renormalized vertex function with $k$ and $p$ the quark four-momenta entering and leaving the vertex and $p-k$ the four-momentum of the gluon fields $A_{\alpha}^{\mu}$ entering the vertex, $\mathfrak{D}'_{\mu\nu}$ is the complete propagator of renormalized gluon fields, and $Z_1$ is the renormalization constant for the quark-gluon vertex function; and that from the added electromagnetic effects
\begin{equation}
i(2\pi)^4\Sigma_e(p)=+Z_2C^2e^2\int\mathrm{d^4}kD'_{\mu\nu}(p-k)\gamma^{\mu}S'(k)\Gamma^{\nu}(k,p),
\end{equation}
where $C_ne$ is the electric charge of a given quark flavour $n$, $D'_{\mu\nu}$ is the complete photon propagator, and $\Gamma^{\nu}$ is the quark-photon vertex function, both of which renormalized. The complete quark propagators in Eqs. (38) and (39) are now understood to include radiations from both electromagnetism and strong interaction, with an overall renormalization constant $Z_2$.

To proceed, the strongly interactive part of the quark self-energy is truncated by the replacement\citep{bloch2002,eich2008,qin2012}
\begin{equation}
Z_1g^2\mathfrak{D}'_{\mu\nu}(k)\mathfrak{\Gamma}^{\nu}(p-k,p)\to 4\pi Z_2^2\alpha^{\mathrm{eff}}_s(k^2)D_{\mu\nu}(k)\gamma^{\nu},
\end{equation}
where $D_{\mu\nu}(k)=\Pi_{\mu\nu}(k)/k^2$ is the bare propagator of gauge fields in Landau gauge with $\Pi^{\mu\nu}(k)=\eta^{\mu\nu}-k^{\mu}k^{\nu}/k^2$, and this way of factorization makes $\alpha^{\mathrm{eff}}_s$, the single unknown function in the truncation, approximately independent of the renormalization point\citep{eich2008ap}. Asymptotically, it has to approach the lowest-order perturbative behaviour of the running coupling constant $\alpha_s\equiv g^2/4\pi$ of quantum chromodynamics,
\begin{equation}
\alpha^{\mathrm{eff}}_s(k^2)\to\frac{12\pi}{(33-2n_f)\mathrm{ln}(k^2/\Lambda^2)}
\end{equation}
for $k^2\to\infty$, where $n_f$ is the number of quark flavours with masses below the energies of interest, and $\Lambda$ is a parameter constant with the dimensions of mass, chosen to make $\alpha_s$ continuous at each quark mass. The infrared behaviour of $\alpha^{\mathrm{eff}}_s$, where perturbation theory fails, requires parameterization. To the same order we consider one-photon exchange in Landau gauge for the electromagnetic part of the quark self-energy:
\begin{equation}
e^2D'_{\mu\nu}(k)\Gamma^{\nu}(p-k,p)\to 4\pi\alpha D_{\mu\nu}(k)\gamma^{\nu},
\end{equation}
where the fine structure constant $\alpha$ is taken approximately as $1/137$, with the understanding that the product $e^2D'_{\mu\nu}(k)$ is renormalization-point independent. Goldstone theorem requires we should use a reciprocal ladder approximation to the Bethe-Salpeter kernel\citep{mun95}
\begin{equation}
i(2\pi)^4 K_1(p,k)\to-\frac{4}{3}[4\pi Z_2^2\alpha^{\mathrm{eff}}_s((p-k)^2)D_{\mu\nu}(p-k)]\gamma^{\mu}\otimes\gamma^{\nu},
\end{equation}
which is consistent with the rainbow truncation to the quark self-energy, Eq. (40); and (42) leads to
\begin{equation}
i(2\pi)^4 K_2(p,k)\to-4\pi Z_2C_nC_m\alpha D_{\mu\nu}(p-k)\gamma^{\mu}\otimes\gamma^{\nu},
\end{equation}
where $C_ne$ and $C_me$ are the corresponding electric quark charges associated with $\chi_{nm}$ in Eq. (36).

In solving the integral Bethe-Salpeter equation, we shall use a matrix representation for the amplitude $\chi$. In the case of some pseudo-Goldstone bosons like $\pi^{\pm,0}$ and $K^{+,0}$, $\chi$ which transforms as a pseudoscalar can be expanded as a sum of terms proportional to the 16 covariant matrices $1$, $\gamma_{\mu}$, $[\gamma_{\mu},\gamma_{\nu}]$, $\gamma_5\gamma_{\mu}$, and $\gamma_5$. The most general pseudoscalar $\chi$ can therefore be written as
\begin{equation}
\chi(p;q)=\gamma_5E+\gamma_5\slashed{p}F+\gamma_5\slashed{q}G+[\gamma_{\mu},\gamma_{\nu}]\epsilon^{\mu\nu\xi\rho}p_{\xi}q_{\rho}H,
\end{equation}
with the 'Levi-Civita tensor' $\epsilon^{\mu\nu\xi\rho}$ defined as a totally antisymmetric quantity with $\epsilon^{0123}=1$, where the coefficients $E$, $F$, $G$, and $H$ are functions of the scalar variables in the problem $p^2$ and $p\cdot q$, and the four-momentum $q^{\mu}$ is fixed at the value $q^2=-m^2$ with $m$ the mass of the one-pseudoscalar-meson state.
We may define the Bethe-Salpeter amplitude $\chi^c$ for emitting an anti-peudoscalar meson $B^c$ of four-momentum $q^{\mu}$ by
\begin{align}
\int\mathrm{d^4}y\mathrm{d^4}z& e^{-iq\cdot(c_1y+c_2z)}e^{-ik\cdot y}e^{+ip\cdot z}\langle\mathrm{VAC}|T\{q_{n}(y)\bar{q}_m(z)\}|B^c,\mathbf{q}\rangle\nonumber\\
&\equiv -(2\pi)^4\delta^4(k-p)\chi^c_{nm}(p;q)/[(2\pi)^{3/2}(2q^0)^{1/2}],
\end{align}
where for external fermion lines the incoming relative momentum $k=q_2-c_1q$ and outgoing relative momentum $p=q_3+c_2q$, with $c_1+c_2=1$. Note that the relation between $\chi^c$ and $\chi$ defined by Eq. (21) is given by 
\begin{equation}
\mathscr{C}\chi(p;q)\mathscr{C}^{-1}=\zeta_B\chi^{c}(-p;q)^{T},
\end{equation}
where $\mathscr{C}\equiv\gamma_2\beta$, and the phase factor $\zeta_B$ is the charge-conjugation parity of the particle $B$. For completely neutral particles like $\pi^0$ with $\zeta_{\pi^0}=+1$, $E$, $G$ and $H$ are even functions of $p\cdot q$ while $F$ is an odd function of $p\cdot q$.

The above formalism can be in principle easily extended to ordinary mesons of arbitrary type belonging to other irreducible representations of the Lorentz group. For example, the Bethe-Salpeter amplitude describing a meson like the $K^{\ast +,0}$, $\rho^{\pm ,0}$, $\omega$, $\overline{K^{\ast}}^{-,0}$ of spin 1 which transforms as a four-vector is defined by
\begin{align}
\int&\mathrm{d^4}y\mathrm{d^4}z e^{-iq\cdot(c_1y+c_2z)}e^{-ik\cdot y}e^{+ip\cdot z}\langle\mathrm{VAC}|T\{q_{n}(y)\bar{q}_m(z)\}|\mathbf{q},\sigma\rangle\nonumber\\
&\equiv (2\pi)^4\delta^4(k-p)\chi^{\mu}_{nm}(p;q)[(2\pi)^{-3/2}(2q^0)^{-1/2}e_{\mu}(\mathbf{q},\sigma)],
\end{align}
where $|\mathbf{q},\sigma\rangle$ is a free-particle state vector of three-momentum $\mathbf{q}$ and spin three-component $\sigma=-1$, $0$, or $+1$. The most general four-vector $\chi^{\mu}$ can be expressed as a decomposition into twelve independent Lorentz covariants. Because of the condition imposed upon the polarization vector $e^{\mu}(\mathbf{q},\sigma)q_{\mu}=0$, $\chi^{\mu}$ is orthogonal to $q_{\mu}$ and the number of allowed covariants reduces to $8$, and hence the most general form of $\chi_{\mu}$ may be written as a linear combination of
\begin{equation}
p_{\mu}^{(\mathcal{O})},\ \gamma_{\mu}^{(\mathcal{O})},\ p_{\mu}^{(\mathcal{O})}\slashed{p},\ p_{\mu}^{(\mathcal{O})}\slashed{q},\ [\gamma_{\mu}^{(\mathcal{O})},\slashed{p}],\ [\gamma_{\mu}^{(\mathcal{O})},\slashed{q}],\ p_{\mu}^{(\mathcal{O})}[\slashed{p},\slashed{q}],\ \gamma_5\gamma_{\xi}\epsilon^{\xi\mu\nu\rho}p_{\nu}q_{\rho},
\end{equation}
with the coefficient of each term a function of the scalar variables $p^2$ and $p\cdot q$, where for any four-vector $V_{\mu}$, $V_{\mu}^{(\mathcal{O})}$ is defined as the component of $V_{\mu}$ orthogonal to the four-momentum $q^{\mu}$: $V_{\mu}^{(\mathcal{O})}\equiv \Pi_{\mu\nu}(q)V^{\nu}$.
The charge-conjugate Bethe-Salpeter amplitudes $\chi^c$ for vector mesons may be defined by
\begin{align}
\int&\mathrm{d^4}y\mathrm{d^4}z e^{-iq\cdot(c_1y+c_2z)}e^{-ik\cdot y}e^{+ip\cdot z}\langle\mathrm{VAC}|T\{q_{n}(y)\bar{q}_m(z)\}|\mathbf{q},\sigma,c\rangle\nonumber\\
&\equiv -(2\pi)^4\delta^4(k-p)\chi^{c}_{\mu nm}(p;q)[(2\pi)^{-3/2}(2q^0)^{-1/2}e^{\mu}(\mathbf{q},\sigma)],
\end{align}
from which we can get easily get the relation between $\chi^c$ and $\chi$ defined by Eq. (48) just as Eq. (47), with $\zeta$ now the charge-cojugation parities of the vector mesons. For particle species that are their own antiparticles like $\rho^0$ with $\zeta_{\rho^0}=-1$, the scalar functions for the Dirac covariants $p_{\mu}^{(\mathcal{O})}\slashed{q}$ and $[\gamma_{\mu}^{(\mathcal{O})},\slashed{p}]$ are odd functions of $p\cdot q$, while those for all the other covariants in Eq. (49) are even functions of $p\cdot q$.

\section{Numerical Results}
To evaluate the integral over some $k$ as for instance in the Bethe-Salpeter equation (36), we perform the Wick rotation of the $k^0$ contour of integration so that in effect $k^0$ is replaced with $ik^4$, with $k^4$ running from $-\infty$ to $+\infty$. This is equivalent to say we now work in Euclidean spacetime, to gain some insight of the effect of electromagnetism in bound states of quark-antiquark pairs by numerical justification.

The angular dependence of the scalar functions of the Bethe-Salpeter amplitudes as in Eq. (45) can be expanded in the forms of
\begin{equation}
E(p^2, p\cdot q; q^2)=\sum_{n=0}\tilde{E}_n(p^2; q^2)U_n(\tau),
\end{equation}
and similar expansions for $F$, $G$, $H$ in (45) and other scalar functions for the Lorentz covariants in Eq. (49), where the real variable $\tau\equiv p\cdot q/(\sqrt{p^2}\sqrt{q^2})$ with $\sqrt{p^2}$ and $\sqrt{q^2}$ the magnitudes of the four-momenta $p^{\mu}$ and $q^{\mu}$,\footnote{The magnitude of the total four-momentum $q^{\mu}$ of the quark-antiquark bound state is purely imaginary in Euclidean spacetime, because of the constraint $q^2=-m^2$ where $m$ is the mass of the bound state.} and $U_n(\tau)$ is the Chebyshev polynomial of the second kind of degree $n$ with $\tau$ in the interval $[-1,1]$, which satisfies a continuous orthogonality relation:
\begin{equation}
\int_{-1}^{1}\mathrm{d}\tau\sqrt{1-\tau^2}U_n(\tau)U_m(\tau)=\frac{\pi}{2}\delta_{nm}.
\end{equation}
Note that the first few Chebyshev polynomials of the second kind are
\begin{equation}
U_0(\tau)=1,\quad U_1(\tau)=2\tau,\quad U_2(\tau)=4\tau^2-1,\quad U_3(\tau)=8\tau^3-4\tau,\quad\dots.
\end{equation}
The dominent contributions to our calculation for the vector and peudoscalar mesons come from the first one or two terms as in the expansion (51); the corrections due to terms of next orders in the Chebyshev expansion are presupposed to be small.\citep{nieu96} The Bethe-Salpeter equations are to be solved numerically by using standard precedures\citep{har96} for the projected amplitudes like the $\tilde{E}(p^2; q^2)$ in Eq. (51) in addition to the spectrum of mesons of interest. The computer memory required to restore and compute digital data can be dramatically reduced by the method of Chebyshev expansion.

As for the effective coupling of the strong interaction $\alpha_s^{\mathrm{eff}}$ in Eq. (40) that will be substituted in Eq. (43), it is understood that it should exhibit sufficient strength for very soft gluons to enable dynamical breaking of chiral symmetry,\citep{mr97} which translates into noble non-perturbative enhancement of the quark dressing functions $A(p^2)$ and $M(p^2)\equiv B(p^2)/A(p^2)$ in Eq. (31) at small momentum $p$. Several models for $\alpha^{\mathrm{eff}}_s$ combining the ultraviolet behaviour as in Eq. (41) with an \emph{Ansatz} in the infrared have been proposed in the past and applied to meson studies.\citep{frank96,mr97,alko2002} In our present work we mainly employ the Maris-Tandy effective interaction\citep{mt99}, which displays the infrared enhancement as a finite-width approximation to $\delta^4(k)$ and the asymptotic part in a form which is deprived of singularities for all real $k^2$,
\begin{equation}
\frac{\alpha^{\mathrm{eff}}_s(k^2)}{k^2}=\frac{\pi\mathscr{D}}{\omega^6}k^2e^{-k^2/\omega^2}+\frac{12\pi}{(33-2n_f)}\frac{2\mathscr{F}(k^2)}{\mathrm{ln}(e^2-1+(1+k^2/\Lambda^2)^2)},
\end{equation}
where $\mathscr{F}(k^2)=(1-\mathrm{exp}(-k^2/4m_t^2))/k^2$, with $m_t=0.5$ GeV interpreted as the mass scale in this model that marks the transition from the perturbative and non-perturbative regime, and $n_f=4$ for which $\Lambda=234$ MeV.
For the quark propagator of the form Eq. (31) that will be used in the Bethe-Salpeter integral equation (36) where the 'center-of-mass' parameters may be chosen as $c_1=c_2=1/2$, we adopt a mass-independent renormalization prescription
\begin{equation}
A(\mu^2)=1, \qquad B(\mu^2)=m_{\mu},
\end{equation}
at some mass scale $\mu=19$ GeV, which is sufficiently large to be in the perturbative domain. Various parameter sets characterized by different values of $\omega$ in Eq. (54), for each of which $\mathscr{D}$ is treated as a parameter of phenomenology to be fitted, along with the renormalized quark masses in Eq. (55), can be given to obtain a good description of the masses of the pion and kaon. Then the masses of heavier mesons like those of vector types and all the associated factors involved in meson decay processes via electroweak interactions can be studied without parameter adjustment. Remember the decay constants for pseudoscalar mesons are defined by Eq. (18) and expressed in terms of Bethe-Salpeter amplitudes by Eq. (23). The decay constants for vector mesons can be defined and expressed just similarly to the pseudoscalar case.\citep{gass82,ivanov99}

With no electromagnetic interactions included, in the original paper of Ref. [23] three different parameter sets have been considered to the extent that the $SU(2)$ isotopic spin invariance is respected, fitted to give a good description of $m_{\pi/K}$ and $F_{\pi}$, and the values for $F_K$ and the masses and decay constants of the vector mesons are calculated afterwards using the obtained fitted values. In Table I we have summerized the results for one parameter set neglecting the effect of electromagnetism, as a numerical check of the Chebyshev expansion method and our Lorentz decomposions of Bethe-Salpeter amplitudes in Eq. (45) and for (49).

We begin to take into account of the eletromagnetic corrections by incorporating kernels in Eqs. (44) and (39) for the integral equations for the Bethe-Salpeter amplitudes as well as the quark propagators, using existed parameter values of the strong effective interaction $\alpha_s^{\mathrm{eff}}$ and fit the degenerate masses of u and d quarks to give a pion mass equal to $m_{\pi^0}=135$ MeV, and then calculate the electromagnetic splitting of the pion mass and also $F_{\pi^{\pm,0}}$, standing upon the observation that from Eq. (35) we see the neutral pion is the true Goldstone boson in the limit of zero quark masses. Those results for three different parameter sets of the Maris-Tandy \emph{Ansatz} (54) are shown in Table II, together with the results accomplished using one intermediate parameter set of the Qin-Chang model\citep{qin2011}. Such a calculation tests the predictive power of the methods used and of the assumptions made about the effective interaction $\alpha_s^{\mathrm{eff}}$.

We are now in a good position to split the quark masses $m_u$ and $m_d$ and get a reasonable estimate of the ratios among $m_u$, $m_d$ and $m_s$. From the mass formulae (35) we may derive formulae for the quark mass ratios in terms of the pion and kaon masses:\citep{wein77}
\begin{align}
&\frac{m_u}{m_d}=\frac{2m_{\pi^0}^2-m_{K^0}^2-m_{\pi^+}^2+m_{K^+}^2}{m_{K^0}^2+m_{\pi^+}^2-m_{K^+}^2},\nonumber\\
&\frac{m_s}{m_d}=\frac{m_{K^0}^2-m_{\pi^+}^2+m_{K^+}^2}{m_{K^0}^2+m_{\pi^+}^2-m_{K^+}^2}.
\end{align}
Using experimental values of the pion and kaon masses gives the ratios $m_u/m_d=0.561$ and $m_s/m_d=20.21$. Thus the ratio of the masses of the d and u quarks is closer to 2 than 1, and the s quark mass is much larger than d and u quark masses.
In the present Bethe-Salpeter formalism we adapt the quark masses as well as their ratios by treating them as input parameters for the correct experimental data of masses of psedoscalar mesons. Indeed the pion isospin multiplet is the one whose mass difference has been successfully calculated based upon one-photon exchange alone,\citep{das67} and hence we determine the numerical values of the u and d quark masses using the one-photon exchange kernel in Eqs. (42) and (44) by matching with the experimental measurements of $m_{\pi^0}$ and $m_{\pi^{\pm}}$; and fit the s quark mass to give consistent values of $m_{K^0}$ and $m_{K^{\pm}}$, since $m_s$ is to be considerably larger than $m_u$ and $m_d$. Then we use the chosen quark masses to obtain results of leptonic decay constants and the properties of vector mesons. In Table III only those values are quoted for which there are reliable experimental data.

\section{Conclusions}
With the electromagnetic interaction treated as an additive correction to the Lagrangian of quantum chromodynamics, the new vertex with one incoming photon line, one incoming quark line and one outgoing quark line manifests itself in the generalized axial-vector Ward identity. The vacuum expectation value of the current associated with the new vertex has a pole whose residue may be expressed in terms of the Bethe-Salpeter amplitude which gives the sum of all Feynman diagrams with external lines of photon and quark fields. The axial-vector Ward identity in this case entails a relationship between this residue and those from the pole contributions of the matrix elements of the axial-vector and ordinary pseudoscalar currents between vacuum states, which is valid to all orders in perturbation theory. The virgin identity can recover to first order in quark masses the expression obtained by old-fashioned effective field studies.

Electromagnetism is meaningful and necessary in the calculation of the spectrum of light mesons of s-wave quark-antiquark bound states in the Bethe-Salpeter formalism. The observed mass difference between the charged and neutral pions turns out to be in greater part from electromagnetism, independent of the assumptions about the form of the quark-antiquark scattering kernel. Only with electromagnetism taken into account, are we able to obtain the values of the individual quark masses and their ratios from the values of the pion and kaon masses. Compared with other pieces of information, however, estimates of the proper mass values of quarks are less reliable than those of the mass ratios; and indeed the definte values for the quark masses given in this article is dependent upon the renormalization prescription adopted. We have used an existent effective quark-antiquark interaction function $\alpha_s^{\mathrm{eff}}$, with the quark masses fitted to reproduce the pion and kaon masses; the subsequent calculation on the basis of one-photon exchange of the vector meson masses and the decay constants are in good agreement with their experimental values.

\begin{table*}[!htbp]
\caption{Overview of the calculated results of the masses of some ground-state mesons and their leptonic decay constants, using the Maris-Tandy model with the parameters $\omega=0.4$ GeV, $D=0.93$ $\mathrm{GeV}^2$. Leading and subleading Chebyshev projections are included for the psudoscalar mesons while only leading Chebyshev moment are used in the vector case. The quark masses are renormalized by a mass-independent scheme at the point $\mu=19$ GeV. Dimensional quantities are reported in MeV except for the quark mass functions M's which are given in GeV.}
\begin{tabular}{c c c c c c c c c c c c c c c c}
\hline\hline
$m_{u/d}$ & $m_s$ & $A_{u/d}(0)$ & $M_{u/d}(0)$ & $A_{s}(0)$ & $M_{s}(0)$ & $m_{\pi}$ & $F_{\pi}$ & $m_K$ & $F_K$ & $m_{\rho}$ & $F_{\rho}$ & $m_{\phi}$ & $F_{\phi}$ & $m_{K^{\ast}}$ & $F_{K^{\ast}}$\\
\hline
3.73 & 85 & 1.57 & 0.49 & 1.59 & 0.68 & 138 & 131.7 & 497 & 156.1 & 740 & 215 & 1075 & 263.6 & 899 & 223.6\\
\hline\hline
\end{tabular}
\end{table*}

\begin{table}[!hbtp]
\caption{Comparison of the properties of the pion, for the three different parameter sets of the effective interaction (54), and also for the parameterization of Qin-Chang \emph{Ansatz}, Ref. [26]. The degenerate quark mass is renormalized by the prescription (55) at the point $\mu=19$ GeV. All the quantities here with the dimensions of mass are released in MeV.}
\begin{tabular}{c c c c c}
\hline\hline
& Ref. [26] &\multicolumn{3}{c}{Maris-Tandy Model}\\
& $\omega=0.5$ GeV & $\omega=0.3$ GeV & $\omega=0.4$ GeV & $\omega=0.5$ GeV\\
& $D=1.02$ $\mathrm{GeV}^2$ & $D=1.25$ $\mathrm{GeV}^2$ & $D=0.93$ $\mathrm{GeV}^2$ & $D=0.79$ $\mathrm{GeV}^2$\\
\hline
$m_{u/d}$ & 3.48  & 3.55 & 3.57 & 3.44\\
$F_{\pi^0}$ & 131.86   & 131.3 & 131.59  & 131.05\\
$m_{\pi^+}$ &  137.9 & 137.8 & 137.9 & 138.1\\
$F_{\pi^+}$ & 131.1 & 131.47 & 131.59 & 130.87\\
$m_{\pi^+}-m_{\pi^0}$ & 2.9 & 2.8 & 2.9 & 3.1\\
\hline\hline
\end{tabular}
\end{table}

\begin{table}[!htbp]
\caption{Calculated values of the masses of the quarks and mesons and the meson decay constants, using one-photon exchange kernel and a reasonable effective quark-antiquark interaction (54) with parameter set $\omega=0.4$ GeV, $D=0.93$ $\mathrm{GeV}^2$. Experimental data are all taken from 'Review of Particle Physics'\citep{pdg2012}, where the quark masses are estimates in a mass-independent subtraction scheme such as $\overline{\mathrm{MS}}$ at a scale $\mu\approx 2$ GeV, while our quark masses evaluated in the Bethe-Salpeter framework are renormalized by the mass-independent strategy (55) at the point $\mu=19$ GeV. Dimensional quantities are broadcasted in MeV.}
\begin{tabular}{c c c}
\hline\hline
& Expt. & Theory\\
& (est.) & (model-dependent)\\
\hline
$m_u/m_d$ & 0.38-0.58 & 0.519\\
$m_s/m_d$ & 17-22 & 17.43 \\
$m_u$ & $2.3^{+0.7}_{-0.5}$ & 2.50\\
$m_d$ & $4.8^{+0.5}_{-0.3}$ & 4.82\\
$m_s$ & $95\pm 5$ & 84\\
$m_{\pi^+}-m_{\pi^0}$ & 4.59 & 4.57\\
$m_{\pi^0}$ & 134.98 & 135.0\\
$m_{\pi^+}$ & 139.57 & 139.6\\
$m_{K^0}-m_{K^+}$ & 3.9 & 3.9\\
$m_{K^0}$ & 497.6 & 497.6\\
$m_{K^+}$ & 493.7 & 493.7\\
$F_{K}/F_{\pi}$ & 1.20 & 1.18\\
$F_{\pi}$ & 130.4 & 131.7\\
$F_{K}$ & 156 & 155.2\\
$m_{K^{\ast 0}}-m_{K^{\ast +}}$ & $6.7\pm 1.2$ &4.8\\
$m_{K^{\ast 0}}$ & 895.8 & 899.9\\
$m_{K^{\ast +}}$ & 891.7 & 895.1\\
$F_{K{\ast}}$ & 225 & 236.1\\
$m_{\rho}$ & 775.3 & 740\\
$F_{\rho}$ & 216 & 214.3\\
\hline\hline
\end{tabular}
\end{table}

\begin{acknowledgments}
We would like to thank Adnan Bashir, Craig Roberts and Wei-Min Sun for helpful discussions. We wishes also to thank Peter Tandy for sharing his manuscripts with us. 
\end{acknowledgments}

\bibliography{em}

\end{document}